\documentclass[conference]{IEEEtran}
\IEEEoverridecommandlockouts

\usepackage{natbib}        
\usepackage{graphicx}      
\usepackage{amsmath,amsfonts} 
\usepackage{url}
\usepackage{array}
\usepackage{enumitem}
\usepackage{xcolor}
\usepackage{eso-pic}

\makeatletter
\let\MYcaption\@makecaption
\makeatother

\usepackage[font=footnotesize]{subcaption}

\makeatletter
\let\@makecaption\MYcaption
\makeatother

\usepackage{pgfplots}
\usepackage{tikz}
\usepackage{color}
\usetikzlibrary{plotmarks,arrows}
\pgfplotsset{compat=1.14}
\graphicspath{{./figures/}{./tikz/figures/}}

\definecolor{mblue}{rgb}{0.00000,0.44700,0.74100}%
\definecolor{morange}{rgb}{0.85000,0.32500,0.09800}%
\definecolor{myellow}{rgb}{0.92900,0.69400,0.12500}%
\definecolor{mpurple}{rgb}{0.49400,0.18400,0.55600}%
\definecolor{mgreen}{rgb}{0.4660    0.6740    0.1880}%
\definecolor{mlightblue}{rgb}{0.3010    0.7450    0.9330}%
\definecolor{mred}{rgb}{ 0.6350    0.0780    0.1840}%
\definecolor{mycolor1}{rgb}{0.00000,0.44700,0.74100}
\definecolor{mycolor2}{rgb}{0.85000,0.32500,0.09800}
\definecolor{mycolor3}{rgb}{0.92900,0.69400,0.12500}
\definecolor{mycolor4}{rgb}{0.49400,0.18400,0.55600}
\definecolor{mycolor5}{rgb}{0.46600,0.67400,0.18800}

\newcommand{\tikzline}[1]{(\protect\tikz[baseline=-0.6ex,x=1pt,y=1pt]{ \protect\draw[#1,thick] [-] (0,0) -- (10,0);})}
\newcommand{\tikzdashedline}[1]{(\protect\tikz[baseline=-0.6ex,x=1pt,y=1pt]{ \protect\draw[#1,thick,dashed] [-] (0,0) -- (10,0);})}
\newcommand{\tikzdashdottedline}[1]{(\protect\tikz[baseline=-0.6ex,x=1pt,y=1pt]{ \protect\draw[#1,thick,dash dot] [-] (0,0) -- (10,0);})}
\newcommand{\tikzdottedline}[1]{(\protect\tikz[baseline=-0.6ex,x=1pt,y=1pt]{ \protect\draw[#1,thick,dotted] [-] (0,0) -- (10,0);})}

\DeclareMathOperator*{\argmin}{arg\,min}

\newcommand{\figref}[1]{\figurename~\ref{#1}}

\newtheorem{remark}{Remark}

\def\BibTeX{{\rm B\kern-.05em{\sc i\kern-.025em b}\kern-.08em
    T\kern-.1667em\lower.7ex\hbox{E}\kern-.125emX}}

\begin{document}
	
	\AddToShipoutPictureBG*{%
		\AtPageUpperLeft{%
			\setlength\unitlength{1in}%
			\hspace*{\dimexpr0.5\paperwidth\relax}
			\makebox(0,-0.75)[c]{
				 \parbox{\paperwidth}{ \centering 
				 Enzo Evers, On Frequency Response Function Identification for Advanced Motion Control, \\
				In {\em IEEE International Workshop on Advanced Motion Control}, Agder, Norway, 2020 } }%
	}}

\title{On Frequency Response Function Identification for Advanced Motion Control
\thanks{This work is supported by the Advanced Thermal Control consortium (ATC), and is part of the research programme VIDI with project number 15698, which is (partly) financed by the Netherlands Organization for Scientific Research (NWO).} }

\author{\IEEEauthorblockN{1\textsuperscript{st} Enzo Evers}
\IEEEauthorblockA{Eindhoven University of Technology \\ Department of Mechanical Engineering \\ Control Systems Technology \\
Eindhoven, The Netherlands. \\ \texttt{e.evers@tue.nl}}
\and
\IEEEauthorblockN{2\textsuperscript{nd} Robbert Voorhoeve}
\IEEEauthorblockA{Eindhoven University of Technology \\ Department of Mechanical Engineering \\ Control Systems Technology \\
	Eindhoven, The Netherlands. \\ \texttt{r.j.voorhoeve@tue.nl}}
\and
\IEEEauthorblockN{3\textsuperscript{rd} Tom Oomen}
\IEEEauthorblockA{Eindhoven University of Technology \\ Department of Mechanical Engineering \\ Control Systems Technology \\
	Eindhoven, The Netherlands. \\ \texttt{t.a.e.oomen@tue.nl}}
}

\maketitle

\begin{abstract}
A key step in control of precision mechatronic systems is Frequency Response Function (FRF) identification. The aim of this paper is to illustrate relevant developments and solutions for FRF identification for advanced motion control. Specifically dealing with transient and/or closed-loop conditions that can normally lead to inaccurate estimation results. This yields essential insights for FRF identification for advanced motion control that are illustrated through a simulation study and validated on an experimental setup. 
\end{abstract}



\section{Introduction}
Many mechatronic systems in the manufacturing industry are considered as Multiple-Input Multiple-Output (MIMO) systems in view of control. These systems often have multiple Degrees of Freedom (DOF) that must be controlled using feedback control for various reasons, e.g., safety margins or constraints on movement range. Furthermore, there is an increasing need for the control of systems-of-systems \citep{evers_beyond_2019} where multiple subsystems jointly contribute to the overall system performance.

Due to the increasing complexity of these MIMO systems-of-systems appropriate modeling techniques are required. For this, acquiring the frequency response function (FRF) of the system is an important first step. FRF identification is often fast and inexpensive and provides an accurate representation of the system. The FRFs can be used for many applications, e.g., direct controller tuning \citep{karimi_robust_2014} or as a basis for parametric modeling \citep{voorhoeve_estimating_2016}. 

The identification of FRFs has made significant progress in recent years, particularly by explicitly addressing
transients errors \citep{schoukens_nonparametric_2009,mckelvey_non-parametric_2012}. Indeed, one of the underlying assumptions is that the system is in steady state, which is often not valid for experimental systems. Furthermore, these approaches have been extended to MIMO systems \citep{voorhoeve_non-parametric_2018}, but in MIMO identification for control, it is often ambiguous which closed-loop transfer functions have to be identified. 

Although important progress is made in FRF identification, application of these advanced methods, especially for multivariable systems, is strikingly limited. The aim of this paper is to provide a clear and concise overview of the steps and decisions that are taken during the identification process. Specifically, two items are investigated in more detail: 1) the elimination of transients and 2) closed-loop aspects for single and multivariable systems. The elimination of transients, i.e. aspect 1), is key when identifying complex systems with many inputs and outputs, commonly, an individual experiment is required for each separate input channel. An approach is presented that can explicitly estimate and remove transient components from FRF estimation that otherwise would cause a biased estimate. Traditionally, these transient effects are mitigated by increasing the experiment length and removal of the initial transient data. By applying the proposed method, a significant reduction in measurement time is achieved. 

Moreover, closed-loop aspects, i.e., aspect 2), are highly important since these increasingly complex systems often operate in closed-loop conditions.Identification under these conditions is increasingly challenging since the additional feed-back loop can cause an estimation bias when not appropriately addressed. Furthermore, a distinction in view of system modeling for control must be made between full MIMO modeling or appropriate selection of a closed-loop Single-Input Single-Output (SISO) transfer function that includes the effect of the interaction between different DOFs. 
\section{Problem formulation} \label{sec:problem}
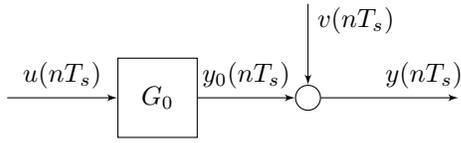
\begin{figure}
\centering
\tikzstyle{block} = [draw, fill=black!0, rectangle,minimum height=3em, minimum width=3em]
\tikzstyle{sum} = [draw, fill=black!0, circle, node distance=1cm]
\tikzstyle{input} = [coordinate]
\tikzstyle{output} = [coordinate]
\tikzstyle{pinstyle} = [pin edge={to-,thick,black}]
\begin{tikzpicture}[auto, node distance=1.25cm,>=latex']
\node [input, name=input] {};
\node [input, above of=input] (input2) {};

\node [block, right of=input, node distance=2cm] (plant) {$G_0$};
\node [sum, right of=plant, node distance=2cm] (sum1) {};
\node [input, above of=plant] (dis) {};

\node [output, right of=sum1,node distance=2cm] (output) {};
\node [input, above of=plant] (input3) {};

\node [input, above of=sum1] (input1) {};

\draw [->] (input) -- node {$u(nT_s)$} (plant);

\draw [->] (plant) -- node {$y_0(nT_s)$} (sum1);
\draw [->] (input1) -- node [pos=0.2] {$v(nT_s)$} (sum1);
\draw [->] (sum1) -- node [near end] {$y(nT_s)$} (output);
\end{tikzpicture}
		\caption{LTI discrete time system in an open-loop setup.} 
		\label{fig:LTI_sysLTI_discrete_time_system}
\end{figure}

Consider the discrete time signal $u(nT_s),\;n=0,\ldots,\,N-1$, where $N$ is the total amount of samples, and its frequency domain representation $U(k)$ obtained by application of the Discrete Fourier Transform (DFT) defined as \citep{pintelon_system_2012}:
\begin{align}\label{eq:DFT}
X(k) &= \dfrac{1}{\sqrt{N}} \sum_{n=0}^{N-1} x(nT_s)e^{-j2\pi nk/N}. 
\end{align}
When the signal $u(nT_s)$ is applied as input to a linear time invariant system $G_0$ with additive output noise $v(nT_s)$, as in \figref{fig:LTI_sysLTI_discrete_time_system}, the resulting output in the frequency domain equals
\begin{equation}
Y(k) = G_0(\Omega_k) U(k) + T(\Omega_k) + V(k)\,,
\label{eq:Output_eq_single}
\end{equation}
where $T(\Omega_k)$ represents the transient contribution and $V(k)$ represents the noise contribution. The argument $\Omega_k$ denotes the generalized frequency variable evaluated at DFT-bin $k$, which, when formulated in, e.g., the Laplace domain, becomes $\Omega_k=j\omega_k$ and in the $Z$-domain $\Omega_k = e^{j\omega_k T_s}$. 

\subsection{Problem formulation}
In this paper focus is placed on two aspects in particular, 1) transient contributions and 2) closed-loop aspects. 
\subsubsection{Transient contribution}

The transient contribution consists of the additional signal that results from past inputs, minus the missed signal in the future response that results from final conditions in the current window. 
Provided that $G_0$ is proper, this yields the following state-space representation of the transient contribution
\begin{equation}
T(z^{-1}) = C(I-z^{-1}A)^{-1}\left(x(0) - x(N)\right). \label{eq:T_states_decaying}
\end{equation}
where the initial state $x(0)$ and final state $x(N)$ capture the past and final conditions respectively.
The additional term $T(\Omega_k)$ in \eqref{eq:Output_eq_single} poses difficulties when identifying the system in transient conditions, i.e., when $x(0) \ne x(N)$. It is generally not possible to separate the forced and transient contribution in the obtained mixed output signal. In Sec. \ref{sec:LPM} a method is provided to alleviate these difficulties. 

\subsubsection{Closed-loop aspects}
Consider again the setup in \figref{fig:LTI_sysLTI_discrete_time_system}, here $u$ is assumed to be independent of $y$ and noise free.
Clearly, in a closed-loop setting, where $u = K(r-y)$, where $K$ is the controller and $r$ and $y$ the reference and output respectively, this no longer holds since $u$ and $y$ are correlated
In view of identification, additional care has to be taken, as is shown in Sec. \ref{sec:CL}.
\subsection{Experimental Setup} \label{sec:exp_setup}
\begin{figure}[ht!]
	\centering			
	\begin{subfigure}[b]{\columnwidth}
		\centering
		\setlength{\fboxsep}{0pt}
		\fbox{\includegraphics[width=0.65\linewidth]{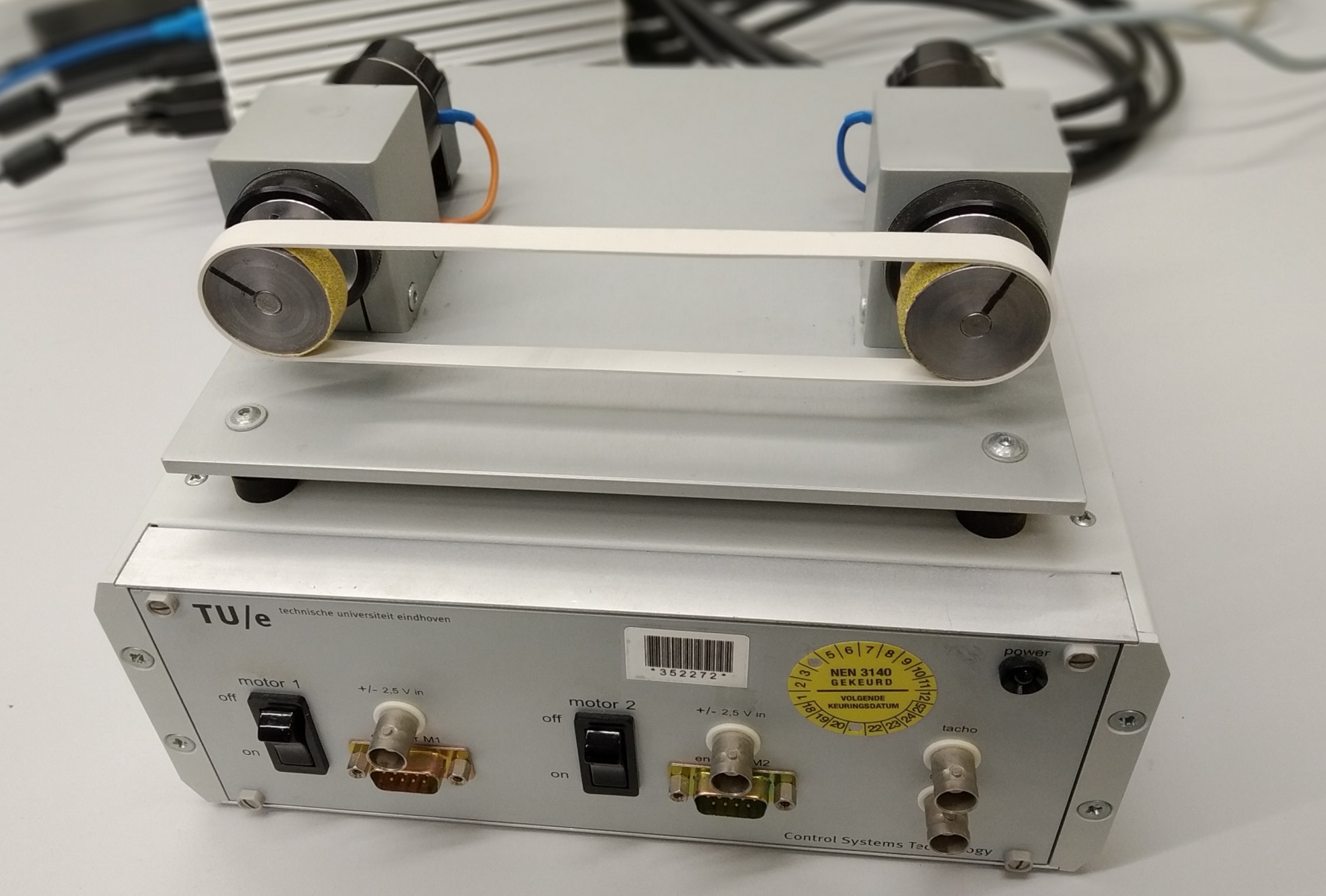}}
		\caption{}%
		\label{fig:exp_setup1}\hfill%
	\end{subfigure} \\
	\begin{subfigure}[b]{\columnwidth}
		\centering
		\includegraphics[width=0.75\linewidth]{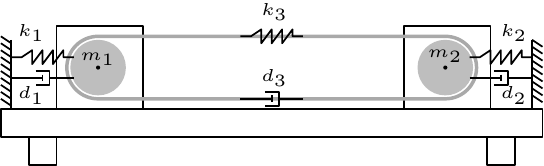}
		\caption{}%
		\label{fig:exp_setup2}\hfill%
	\end{subfigure} 
	\caption{The experimental setup (a) and it's schematic counterpart (b). The system consists of two DC motors coupled by an elastic connection, often representative for a conveyor belt type system. Due to the direct connection between the motors, the system is highly multivariable.}
	\label{fig:exp_setup}
\end{figure}

The challenges for FRF identification presented in this paper are demonstrated on an experimental setup or a representative simulation model. 
The experimental setup is shown in Fig. \ref{fig:exp_setup} and it consists out of two DC motors interconnected by a flexible connection. The system is operated in closed-loop and is multivariable with high interaction terms, i.e., with strong cross-coupling between the two inputs and outputs.

A representative simulation model is constructed by considering the setup as two masses connected to each other and the fixed world by spring-damper elements, as shown in Fig. \ref{fig:exp_setup2}. 
The state-space model of the system is then given by 

{ \small
	\setlength{\abovedisplayskip}{5pt}
	\setlength{\belowdisplayskip}{5pt}
\begin{align} 
\dot{x} &= \underbrace{\begin{bmatrix}
0 & 1 & 0 & 0 \\
-173 & -8 & 166 & 1.33 \\
0 & 0 & 0 & 1 \\
166 & 1.33 & -173 & -8
\end{bmatrix}}_{A} \begin{bmatrix}
x_1 \\ x_2 \\ x_3 \\ x_4
\end{bmatrix} + \underbrace{\begin{bmatrix}
0 & 0 \\ 53 & 0 \\ 0 & 0 \\ 0 & 53
\end{bmatrix}}_{B}\begin{bmatrix}
u_1 \\ u_2
\end{bmatrix} \\
y &= \underbrace{\begin{bmatrix}
1 & 0 & 0 & 0 \\ 0 & 0 & 1 & 0
\end{bmatrix}}_{C} \begin{bmatrix}
x_1 \\ x_2 \\ x_3 \\ x_4
\end{bmatrix}
\end{align} }
where $y$ [rad] is the angular position measured by optical encoders and $u$ [v] is the input voltage to the linear amplifiers that control the motors. 
The transfer function matrix (TFM) can then by obtained by $G(s) = C(sI - A)^{-1}B$.

\section{Estimators} \label{sec:estimators}
In this section, various estimators for Frequency Response Functions (FRFs are presented.
Throughout, the SISO open-loop case is considered. The extension to closed-loop and MIMO is considered in Sec. \ref{sec:CL} and Sec. \ref{sec:CL_mimo} respectively.
\subsection{Empirical Transfer Function Estimate}
One of the most straightforward transfer function estimates can be obtained by simply dividing the frequency domain output signal with the input signal. This is known as the Empirical Transfer Function Estimate or ETFE, i.e.,
\begin{equation}
\hat{G}_\mathrm{ETFE}(k) = Y(k) / U(k)\,. \label{eq:ETFE}
\end{equation}
To average out the noise contribution, the input and output signals are divided into windows of equal length and subsequently the ETFE is obtained for each window $m$ to yield,
\begin{equation}
\hat{G}_\mathrm{ETFE}(k) = \dfrac{1}{M} \sum_m Y_m(k) / U_m(k)\,.
\end{equation}
When periodic excitation signals are used and the window length matches the periodicity of the excitation, this is an effective approach. However when arbitrary excitation signals are used such as noise excitation, averaging as performed in the ETFE can lead to poor results. This is due to the fact that when the excitation signal $u(nT_s)$ is a Gaussian (pseudo) random signal, the amplitude of $U(k)$ is also stochastic and can therefore be arbitrarily close to zero. When $U_m(k)$ is close to singular in some window $m$ and at frequency bin $k$, the ETFE at that frequency will be dominated by the noise contribution yielding a very poor FRF estimate. This poor FRF estimate is subsequently averaged with the estimates in the other windows without accounting for the difference in the quality of the individual estimates.

The main guideline to follow is to always average out the noise before division. When this principle is directly applied for arbitrary signals by first averaging the inputs and outputs over all windows and then performing the division, i.e.,
\begin{equation}
\hat{G}_\mathrm{avg} = Y_\mathrm{avg}(k) / U_\mathrm{avg}(k),
\end{equation}
with
\begin{equation}
Y_\mathrm{avg}(k) = \dfrac{1}{M} \sum_m Y_m(k), U_\mathrm{avg}(k) = \dfrac{1}{M} \sum_m U_m(k),
\end{equation}
another problem occurs due to the randomness of the phases of $U_m(k)$ for each $m$. As these phases are random the mean value, i.e., $U_\mathrm{avg}(k)$, will tend to zero for large $M$.

\subsection{Spectral Analysis} \label{sec:Estimators_SA}
For arbitrary signals the spectral analysis approach is commonly applied. In this approach the cross-power spectrum of the input and output signals and auto-power spectrum of the input signal are first calculated, involving an averaging step over all considered excitation windows, and subsequently the transfer function estimate is obtained by dividing these spectra
\begin{equation}
\hat{G}_\mathrm{SA}(k) = \hat{\Phi}_{yu}(k) / \hat{\Phi}_{uu}(k) \label{eq:estimator_SA}
\end{equation}
where $G_{SA}$ denotes the spectral analysis approach and
\begin{equation}
\hat{\Phi}_{yu}(k) = \dfrac{1}{M} \sum_m Y_m(k) U_m(k)^H
\end{equation}
\begin{equation}
\hat{\Phi}_{uu}(k) = \dfrac{1}{M} \sum_m U_m(k) U_m(k)^H.
\end{equation}
In this approach, transient suppression is achieved by using so called windowing functions, such as the Hanning window. For additional details, see \citet[sections 2.2.3 \& 2.6.5]{pintelon_system_2012}.
An estimate for the noise covariance and the covariance on the transfer function estimate are obtained in spectral analysis through \citep[eq. (7-33) \& (7-42)]{pintelon_system_2012}.
\begin{equation}
\sigma^2_v(k) = \dfrac{M}{M-n_u}\left(\hat{\Phi}_{YY}(k) - \hat{\Phi}_{YU}(k)/\hat{\Phi}_{UU}(k)\hat{\Phi}_{YU}^{H}(k)\right)
\end{equation}

\begin{equation}
\sigma^2_{G_\mathrm{SA}}(k) = \dfrac{1}{M}\left(1/\hat{\Phi}_{UU}(k) \cdot C_v(k)\right)
\end{equation}

\begin{remark}
It is of key importance to quantify the uncertainty on any Frequency Response Function estimate that is obtained. When no such quality measure is provided it is impossible to adequately interpret the obtained estimate seriously impacting the usefulness of the obtained estimate. A quality measure that is often used in spectral analysis is the coherence functions.  While this measure is certainly useful, it is used in a rather qualitative way, where a coherence which is close to unity is indicative of a high quality measurement while a coherence value closer to zero is indicative of a poor estimate. {To facilitate the presentation, in this work some results purposefully exclude the variance of the estimate.}
\end{remark}	
	
\begin{figure*}[h]
	\centering
	\includegraphics[width=0.5\linewidth]{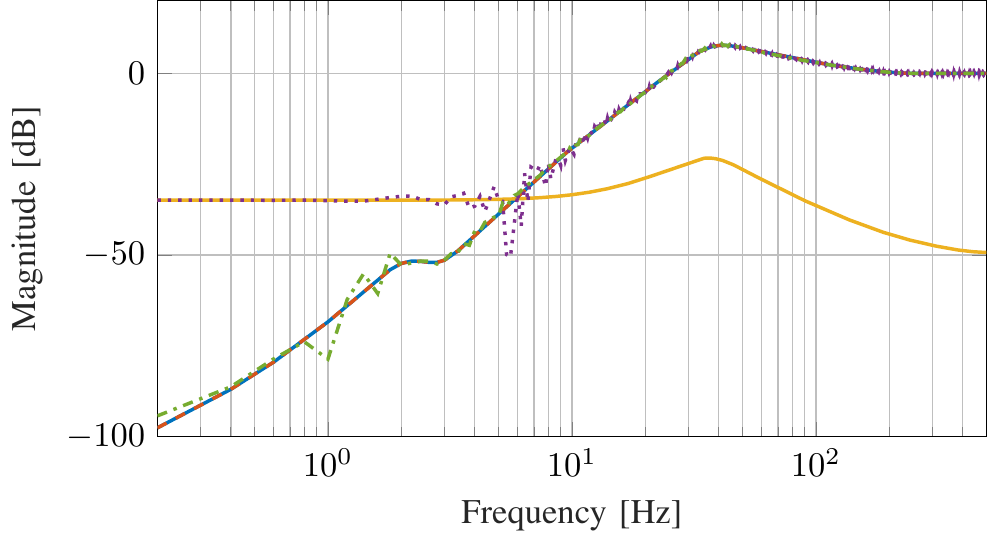}%
	\includegraphics[width=0.5\linewidth]{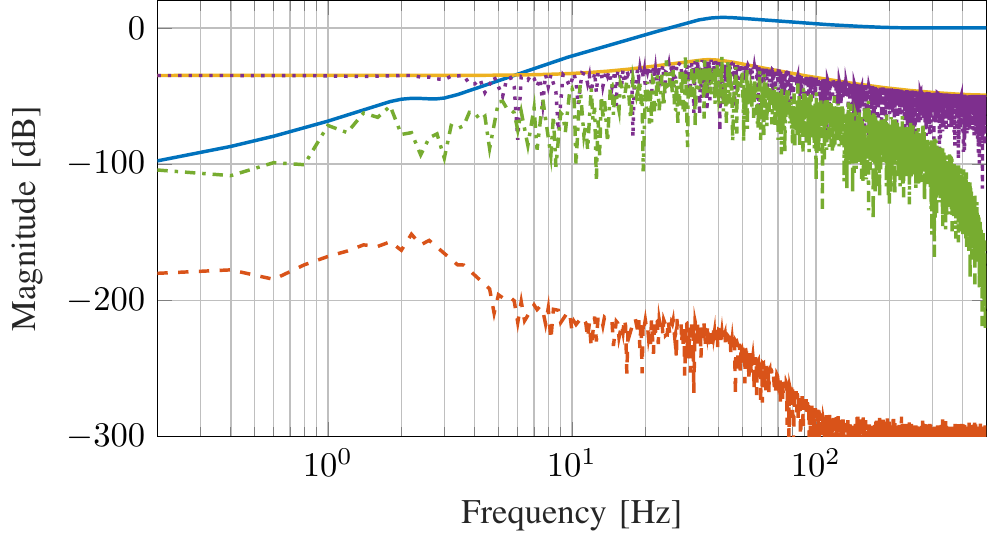}
	\caption{Identification result under transient conditions, the left figure shows the magnitude of the identified frequency response function and the right figure shows its estimation error when compared to the model. The results show that the LPM \tikzdashedline{mycolor2}, described in Sec. \ref{sec:LPM}, outperforms the spectral analysis method using both a rectangular \tikzdottedline{mycolor4} and a Hann \tikzdashdottedline{mycolor5} window. The LPM is invariant to the transient contribution \tikzline{mycolor3} that dominates the response at lower frequencies when compared to the plant \tikzline{mycolor1}. }
	\label{fig:sim_transient}
\end{figure*}

\subsection{Local Modeling Approach} \label{sec:LPM}
The main idea of the local modeling approach, e.g., as in \citet{schoukens_nonparametric_2009}, is to identify a model, with validity over only a small frequency range, which can be used to provide a non-parametric estimate of the FRF and the transient at the central point $k$. Consequently, errors due to the transient can be effectively eliminated. To achieve this, a small frequency window $r$ around DFT-bin $k$ is considered, i.e., $r =[-n_w,\dots,n_w] \in \mathbb{Z}$, to yield
\begin{equation}
Y(k+r) = G(\Omega_{k+r}) U(k+r) + T(\Omega_{k+r}) + V(k+r)\,.
\label{eq:Output_eq_multi}
\end{equation}
Next, both the plant, $G(\Omega_{k+r})$, and the transient contribution, $T(\Omega_{k+r})$, which are assumed to be smooth functions of the frequency, are parametrized. For instance, using the polynomial parametrization
\begin{align}
G(\Omega_{k+r}) &= G(\Omega_k) + \sum_{s=1}^R g_s(k) r^s\,,\label{eq:sys_LPM}\\
T(\Omega_{k+r}) &= T(\Omega_k) + \sum_{s=1}^R t_s(k) r^s\,.\label{eq:trans_LPM}
\end{align}
Using this parametrization \eqref{eq:Output_eq_multi} is rewritten as

\begin{align}
\label{eq:parametrized_OE}
Y(k+r) &=\Theta(k) K(k+r) + V(k+r),
\end{align}
with
\begin{align}
\Theta(k) &= \begin{bmatrix} \Theta_G(k) & \Theta_T(k) \end{bmatrix},\nonumber \\
\Theta_G(k) &= \begin{bmatrix}
G(\Omega_k) & g_1(k) & g_2(k)& \ldots & g_R(k)
\end{bmatrix},\nonumber \\
\Theta_T(k)&= \begin{bmatrix}
T(\Omega_k) & t_1(k) & t_2(k) & \ldots & t_R(k)
\end{bmatrix}, \label{eq:param_LPM}\\
K(k+r) &= \begin{bmatrix}
K_1(r) \otimes U(k+r) \\ K_1(r)
\end{bmatrix}
\end{align}
where \begin{equation}
K_1(r) = \begin{bmatrix}
1&r& \cdots&r^R
\end{bmatrix}^T.
\end{equation}
Finally, the parameters of the local model are determined by solving the linear least squares problem
\begin{align}
\label{eq:LPM_lsq}
\hat{\Theta}(k) &= \argmin\limits_{\Theta(k)}{\sum_{r=-n_W}^{n_W}\left\Vert Y(k+r)-\Theta(k) K(k+r)\right\Vert^2_2} \\ 
				&= Y_n(k) K_n(k)^+\,
\end{align}
with $X_n(k) = \begin{bmatrix}X(k-n_W) &\cdots & X(k+n_W)\end{bmatrix}$, and with $A^+ = A^H(AA^H)^{-1}$ the right Moore-Penrose pseudo-inverse Performing this least squares estimation for each DFT-bin, $k$, and evaluating the local models at the center frequency $r=0$, yields a non-parametric estimate, ${G}(\Omega_k)$, for the FRF. For the parametrization described here this evaluation is trivial, since for $r=0$ only the zeroth order polynomial term remains, which is why ${G}(\Omega_k)$ directly appears in this parametrization, see \eqref{eq:sys_LPM}.

An estimate for the covariance on the transfer function estimate can be obtained as provided in \citet[section 7.2.2.2]{schoukens_frequency_2012}.

\section{Transients in System Identification}
Estimating the true plant $G_0(\Omega_k)$ in \eqref{eq:Output_eq_single} using classical estimators is challenging since the measurement of $Y(k)$ is contaminated with an additional component $T(k)$. In Sec. \ref{sec:LPM} it is shown that an explicit estimation of $T(k)$ can be made by constructing a local parametric model in a frequency window of width $n$ around DFT bin $k$. The underlying assumption facilitating this approach is the smoothness of the transient component $T(k)$ in \eqref{eq:T_states_decaying} since the transient is a decaying response, assuming that the system is stable. 
\paragraph*{Simulation study}
To illustrate the benefit of the local modeling approach, a simulation study is performed using the simulation model presented in Sec. \ref{sec:exp_setup}. The system $G_0$ is considered in closed-loop, see \figref{fig:LTI_sysLTI_discrete_time_system_CL}, with excitation input $d(nT_s)$ and identification output $u(nT_s)$. Therefore, the identification setting essentially is an open-loop one as in \figref{fig:LTI_sysLTI_discrete_time_system}, since the transfer function that is identified is the sensitivity function $S(\Omega_k) = \dfrac{u}{d}$, and $d$ is noise free. 
\paragraph*{Simulation results}
The excitation signal is $2$ periods of a $5$ [s] random phase multisine with a sampling frequency  $F_s = 1000$ [Hz], resulting in a total of $10000$ samples. The FRF of $G_0$ is then estimated using both the spectral analysis and local modeling approach, presented in Sec. \ref{sec:estimators}, yielding results as shown in \figref{fig:sim_transient}. The spectral analysis method is applied using both a rectangular window and a Hann window. While the latter appears to achieve improved performance, the results are misleading, with increasing amount of periods, mitigating the effect of transients, the rectangular window will outperform the Hann window. Indeed, with a periodic excitation no window should be used since this will introduce additional leakage errors. The local modeling approach clearly outperforms the spectral analysis approach, since it explicitly estimates and removes the transient component $T(\Omega_k)$. 
\section{Closed-loop aspects} \label{sec:CL}
\begin{figure}[t]
	\centering
	\includegraphics[width=0.9\linewidth]{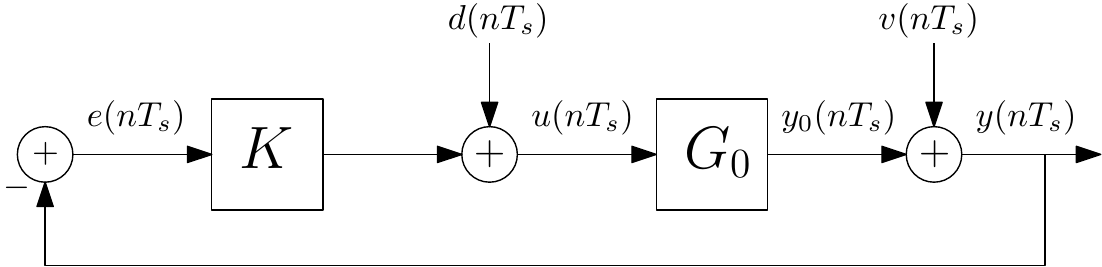}
	\caption{LTI discrete time system in a closed-loop setup.} 
	\label{fig:LTI_sysLTI_discrete_time_system_CL}
\end{figure}

The techniques presented in the previous sections are described for an open-loop output error setup, see, e.g., Fig. \ref{fig:LTI_sysLTI_discrete_time_system}.
Many precision motion systems have safety constraints, requiring closed-loop operations.
In this section, the transition to a closed-loop identification setting is made as shown in Fig. \ref{fig:LTI_sysLTI_discrete_time_system_CL}.
It is shown that the techniques presented in Sec. \ref{sec:estimators} can be straightforwardly extended to a closed-loop setting when taking into account some important differences.
\subsection{Indirect approach to Closed-loop identification} \label{sec:cl_indirect}
A common approach for the control of motion systems is to neglect possible cross-coupling between the DOFs or performing a decoupling procedure.
This allows the simplification to a Single-Input Single-Output (SISO) control setting. 
Consider again an LTI discrete time system, now operating in closed-loop as shown in Fig. \ref{fig:LTI_sysLTI_discrete_time_system_CL}.
A transfer function estimation can be performed using excitation input $d(nT_s)$, identification input $u(nT_s)$ and identification output $y(nT_s)$, as described in Sec. \ref{sec:Estimators_SA}, yielding \citep[Chap. 10]{soderstrom_system_1989}

\begin{equation}
\hat{G}(k) = \frac{\hat{\Phi}_{yu}(k)}{\hat{\Phi}_{uu}(k)} = \frac{G_0\Phi_{dd}-K\Phi_{vv}}{\Phi_{dd} + |K|^2\Phi_{vv}}. \label{eq:SA_CL_bias}
\end{equation}
This clearly yields a biased estimate of $G_0$ since $y$ and $v$ are not independent in the closed-loop setting. This shows that taking a direct approach and identifying the system $G_0$ using $y$ and $u$ in a closed-loop setting can lead to an estimation bias. 

\paragraph*{Indirect approach}
To mitigate the estimation bias resulting from a direct approach in Eq. \ref{eq:SA_CL_bias}, an indirect approach is considered that identifies two transfer functions in a Single Input Multiple Output (SIMO) procedure to yield
\begin{equation}
\widehat{GS}(k) = \dfrac{\hat{\Phi}_{yd}}{\hat{\Phi}_{dd}}, \widehat{S}(k) = \dfrac{\hat{\Phi}_{ud}}{\hat{\Phi}_{dd}}
\end{equation}
that are the estimate of the process sensitivity and sensitivity respectively.
The system estimate $\hat{G}_0$ is then obtained by
\begin{equation}
\hat{G}(k) = \dfrac{\widehat{GS}(k)}{\widehat{S}(k)}.
\label{eq:SA_CL_unbiased}
\end{equation}
 Alternatively $\hat{G}(k)$ is obtained by $\hat{G}(k) = \frac{1}{K(k)}(\frac{1}{S(k)} - 1)$ but this requires the controller $K(k)$ to be known exactly, which is often not possible. The estimate \eqref{eq:SA_CL_unbiased} is unbiased in the closed-loop setting since $d$ and $v$ are independent. Note that \eqref{eq:SA_CL_unbiased} recovers \eqref{eq:estimator_SA} in the open-loop setting since then $S(k) = 1$ and $GS(k) = G(k)$. Similarly, the local approach proposed in Sec. \ref{sec:LPM} can be used to identify $GS(k)$ and $S(k)$ in \eqref{eq:SA_CL_unbiased} to yield an unbiased estimate of $\hat{G}(k)$ under transient conditions.

\begin{figure}[h]
	\centering
	\includegraphics[width=1\linewidth]{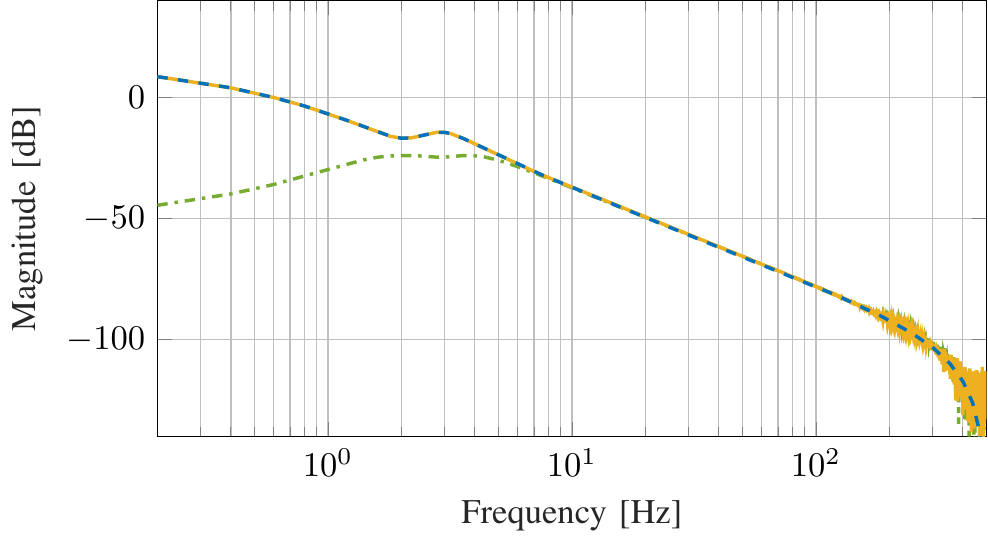}
	\caption{Comparison of the estimated FRF of the true system \tikzdashedline{mycolor1} using the direct method \tikzdashdottedline{mycolor5} and the indirect method \tikzline{mycolor3} . It is shown that applying the direct method in a closed-loop setting yields a significantly biased result.}
	\label{fig:sim_closed_loop_siso}
\end{figure}

\begin{figure}[]
	\centering
	\includegraphics[width=1\linewidth]{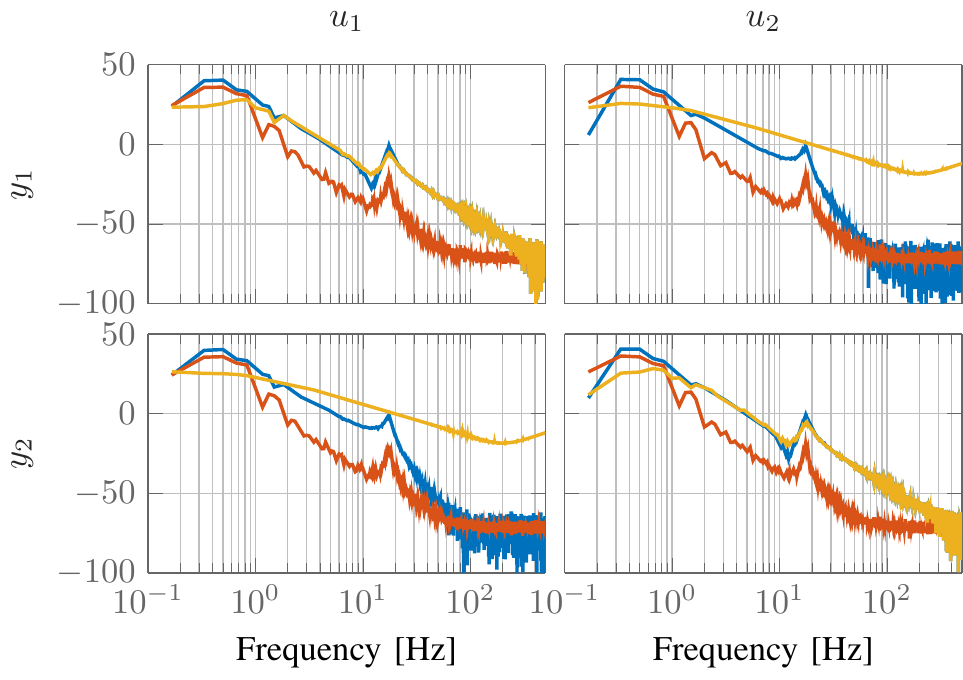}
	\caption{Experimental estimation of the MIMO transfer function matrix, shown as a magnitude [dB] plot, using both matrix wise \tikzline{mycolor1}, with corresponding variance \tikzline{mycolor2}, or element wise division \tikzline{mycolor3}, shown without variance, yielding significantly different models. Depending on the desired model, a specific operator should be used.}
	\label{fig:exp_MIMO_element_vs_matrix}
\end{figure}

\subsection{Closed-loop identification of multivariable systems} \label{sec:CL_mimo}
In this section the indirect method to closed-loop identification is extended to encompass MIMO systems.
Depending on the control objective, a different plant model is desired. A distinction is made between the true multivariable plant and an equivalent plant.

\begin{figure}[h!]
	\centering
	\includegraphics[width=0.9\linewidth]{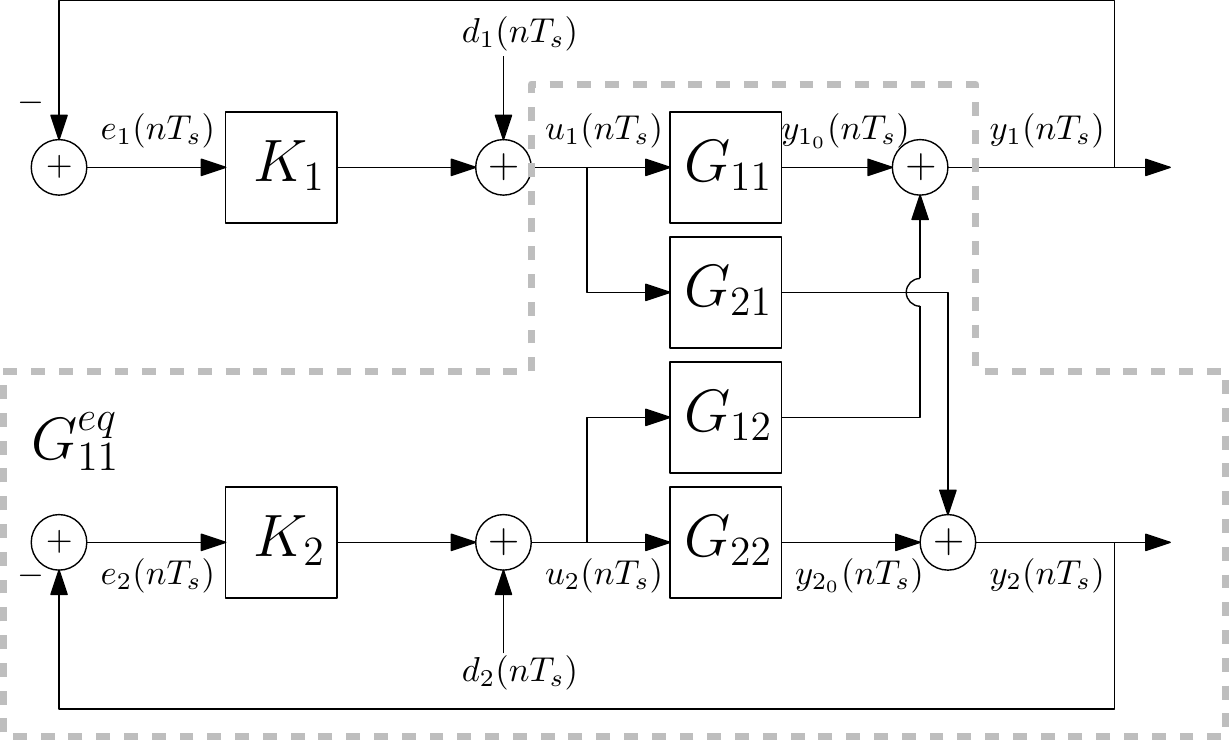}
	\caption{LTI discrete time system in a MIMO closed-loop setup. Here $G^{eq}_{11}$ indicates the equivalent plant that includes the interaction in the secondary loop as a SISO transfer function.} 
	\label{fig:LTI_sysLTI_discrete_time_system_CL_MIMO}
\end{figure}

Consider a MIMO system in the closed-loop setting as shown in Fig. \ref{fig:LTI_sysLTI_discrete_time_system_CL_MIMO}. The system in Fig. \ref{fig:LTI_sysLTI_discrete_time_system_CL_MIMO} is multivariable and the control solution is decentralized, i.e., diagonal. Consider now the proposed indirect method from Sec. \ref{sec:cl_indirect} using excitation input $d_1$ and identification output $u_1,y_1$ and $u_2,y_2$ to identify the first column of $\hat{G}$ to obtain $G_{11},G_{21}$. Applying the indirect method to individual entries of $GS$ and $S$ yields for $G_{11}$

\begin{equation}
\tilde{G}^{eq}_{11} = {GS}_{11}/S_{11} = G_{11} - \underbrace{\frac{G_{12}K_2G_{21}}{1+K_2P_{22}}}_{\mathrm{interaction}}. \label{eq:single_inverse}
\end{equation}
Here, the estimated $\tilde{G}^{eq}_{11}$ is clearly different from the ``true'' MIMO entry $G_{11}$, as indicated in \figref{fig:LTI_sysLTI_discrete_time_system_CL_MIMO}. This is caused by the interaction terms $G_{12},G_{21}$ and the secondary closed-loop controller $K_{22}$. 
Indeed, the plant $\tilde{G}^{eq}_{11}$ is also known as the ``equivalent plant'', since it represents the fully coupled system as a single SISO transfer function. This approach is often applied in sequential loop closing, where a full MIMO system is modeled as a sequence of equivalent plants for which a SISO controller is designed \citep{maciejowski_multivariable_1989}.

To obtain the full multivariable plant $G$ a slightly different formulation to \eqref{eq:single_inverse} is required.
The plant $G$ can straightforwardly be identified using $2$ independent excitations, exciting $d_1$ and $d_2$ to identify the first and second column respectively of $GS$ and $S$. Then $G$ is obtained by performing
\begin{align}
G(\Omega_k) &= GS(\Omega_k)S(\Omega_k)^{-1} 
\end{align}
for each frequency $\Omega_k$. Here, the Frequency Response Matrix (FRM) of the closed-loop sensitivity function $S(\Omega_k)$ is inverted using a matrix inverse and not element wise inversion. While this difference is subtle, the obtained plant is significantly different for systems with interaction.

\section{Experiments}
In this section, the techniques presented in this paper are applied to the experimental setup as shown in Fig. \ref{fig:exp_setup}. Specifically, the transient elimination by employing the local modeling technique, as shown in Sec. \ref{sec:LPM}, and the closed-loop MIMO estimation, as shown in Sec. \ref{sec:CL_mimo}, are highlighted.

\subsection{Transient elimination}
\begin{figure}[]
	\centering
	\includegraphics[width=1\linewidth]{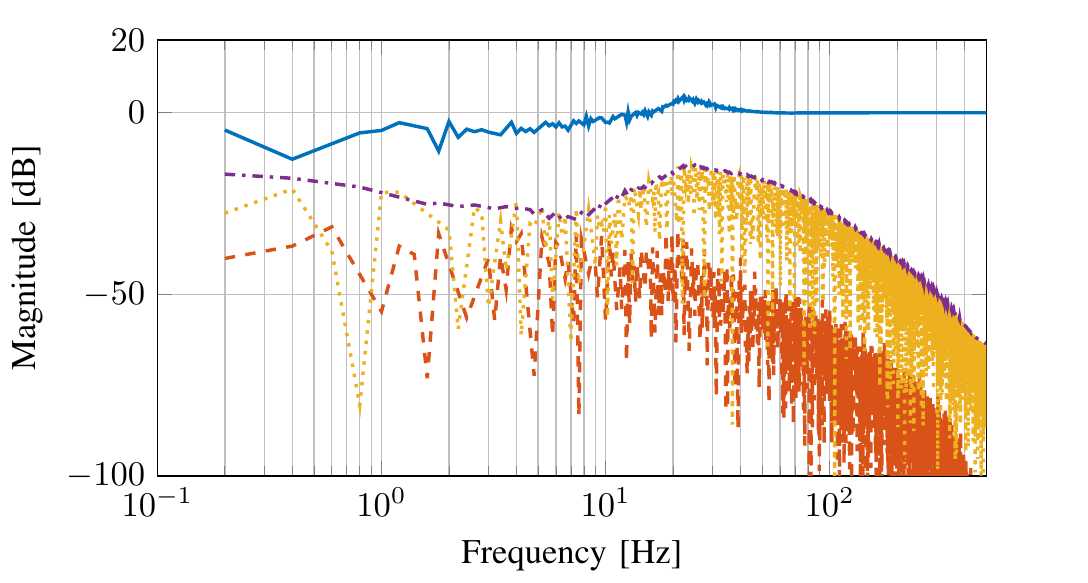}
	\caption{Estimating the FRF of the sensitivity function $S = \frac{u}{d}$ using $2$ periods of a $5$ [s] multisine, compared to using $6$ periods as a baseline reference \tikzline{mycolor1}. Results show that the estimation error using spectral analysis \tikzdottedline{mycolor3} is significantly higher than when using the LPM method \tikzdashedline{mycolor2}, this is caused by the transient contribution \tikzdashdottedline{mycolor4}.}
	\label{fig:exp_transient}
\end{figure}

To illustrate the effects of transient conditions on the estimation accuracy, an FRF is estimated on two different dataset. The first dataset contains $6$ periods of the system response to a multisine signal with a length of $5$ [s], yielding $N = 30 F_s = 30\cdot10^3$ samples, this dataset serves as a baseline reference. The second dataset contains only the first $2$ periods of the first dataset, reducing the number of available samples to $N = 10\cdot F_s = 10\cdot10^3$. Moreover, the initial periods contain significantly more transients than the latter. The results are shown in \figref{fig:exp_transient}. It is demonstrated that the LPM described in Sec. \ref{sec:LPM} is able to significantly reduce the estimation error caused by the transient contributions, when compared to the more classical spectral analysis approach. Moreover, since the LPM can cope with transient data, a significant savings in the required experimental time is achieved.

\subsection{MIMO identification in a closed-loop setting}
In Sec. \ref{sec:CL_mimo} it is shown that by applying the indirect method in multivariable setting, two different FRM can be obtained. By applying the matrix inverse, e.g., $\hat{G} = PSS^{-1}$ the multivariable plant model is obtained. Conversely, if an element wise inversion is employed, e.g., $PS \odot \dfrac{1}{S}$ where $\odot$ is the Hadamard product, then the equivalent plant model is obtained. This is illustrated on the experimental setup as shown in \figref{fig:exp_MIMO_element_vs_matrix}. The results show that depending on the desired model, a different matrix operation should be employed.

\section{Conclusion}
In this paper, an overview of important aspects in FRF identification for advanced motion control, specifically transient and closed-loop conditions, is presented. It is shown that if these aspects are not appropriately addressed the FRF estimate can be biased or of poor quality. By applying the techniques presented in this paper a high quality unbiased FRF is obtained, facilitating parametric modeling or direct controller design for advanced motion control.

\end{document}